\documentclass[12pt,a4paper]{article}
\usepackage[T1]{fontenc}
\usepackage[utf8]{inputenc}
\usepackage{lmodern}
\usepackage[margin=25mm]{geometry}
\usepackage{amsmath,amssymb}
\usepackage{microtype}
\usepackage[hidelinks]{hyperref}
\hypersetup{
  pdftitle={From gauge-covariant Noether invariants to electromagnetic velocity-dependent mechanics},
  pdfauthor={A. S. de Castro}
}
\title{From gauge-covariant Noether invariants to electromagnetic velocity-dependent mechanics}
\author{A. S. de Castro\\[0.6em]
  \small S\~ao Paulo State University (UNESP)\\
  \small School of Engineering and Sciences\\
  \small Campus of Guaratinguet\'a, SP, Brazil}
\date{\normalsize Accepted for publication in \textit{Europhysics Letters}}

\begin{document}
\maketitle

\begin{abstract}
For a charged particle, a spacetime symmetry need preserve the electromagnetic potential only up to a gauge transformation. Starting from the standard Noether--Bessel-Hagen construction, we make explicit the reduction from the relativistic gauge-invariant worldline charge to the generalized constants of nonrelativistic velocity-dependent mechanics, retaining the quasi-invariance boundary term throughout. With the convention $S_p=-mc\int ds+q\int A$, the canonical worldline one-form is $P=-m u^\flat+qA$, and a Killing vector satisfying $\mathcal L_XA=d\chi_X$ generates $C_X=P(X)-q\chi_X$. Its low-velocity limit is the Noether--Bessel-Hagen charge of the coordinate-time Lagrangian, from which generalized linear momentum, angular momentum and energy follow generator by generator. A uniform magnetic field gives an elementary gauge-dependent representative but gauge-invariant translational charge. The Darwin interaction then supplies the conservative $O(c^{-2})$ many-particle realization after field elimination, while a closed particle--field system is governed by the total stress-energy Killing charge. The result is an explicit reduction map linking three established descriptions without identifying their distinct conserved quantities.
\end{abstract}


\section{Introduction}
Electromagnetic interactions provide a basic setting in which spacetime symmetries cannot in general be identified with conservation of mechanical particle momentum alone. Gauge freedom sharpens the issue: canonical generators depend on the potential $A$, whereas the symmetry charge must be independent of the gauge representative. In nonrelativistic mechanics the same structure appears through velocity-dependent interactions $U(\mathbf r_i,\mathbf v_i,t)$, for which the conserved quantities generated by translations, rotations and time translations contain derivatives of $U$ with respect to the velocities~\cite{deCastro2002}.

The relativistic formulation also fixes the scope of the term ``velocity-dependent potential.'' A spacetime one-form $A=A_\mu dx^\mu$ generates the worldline coupling $qA_\mu u^\mu$, which is linear in the four-velocity and describes electromagnetic-type velocity dependence rather than an arbitrary function on the tangent bundle. The relevant symmetry condition is weaker than strict invariance. If $X$ generates a spacetime symmetry, the particle action may be invariant up to a total derivative,
\begin{equation}
 \mathcal L_Xg=0,\qquad \mathcal L_XA=d\chi_X .
 \label{symmetry}
\end{equation}
This is the familiar quasi-invariance underlying the Noether--Bessel-Hagen construction~\cite{Noether1918,BesselHagen1921}.

The ingredients entering the present discussion are individually standard: the gauge-covariant Noether construction, generalized constants for velocity-dependent interactions, and the Darwin approximation are all established. The purpose of this Letter is therefore not to introduce any of them separately. The result established here is the explicit reduction map connecting them while retaining the quasi-invariance boundary term at every stage. In particular, we distinguish (i) the gauge-invariant test-particle worldline charge $C_X$, (ii) its nonrelativistic Noether--Bessel-Hagen descendant $C_X^{\rm NR}$, and (iii) the conservative Darwin many-body charges obtained after field elimination. The closed particle--field stress-energy charge is kept conceptually separate. This organization makes precise which quantities are related by limiting procedures and which belong to different dynamical descriptions~\cite{Blaschke2021,Klinger2024}.

\section{Gauge-covariant particle invariant}
Let $(\mathcal M,g)$ be a Lorentzian spacetime with signature $(+,-,-,-)$. In a local inertial chart we use $x^\mu=(ct,\mathbf r)$ and
\begin{equation}
 A=-\frac{\phi}{c}\,dx^0+\mathbf A\cdot d\mathbf r,\qquad F=dA,
 \label{potential}
\end{equation}
so that $qA_\mu dx^\mu=-q\phi\,dt+q\mathbf A\cdot d\mathbf r$. For a particle of mass $m$ and charge $q$ consider
\begin{equation}
 S_p=-mc\int \sqrt{g_{\mu\nu}\dot x^\mu\dot x^\nu}\,d\lambda
      +q\int A_\mu\dot x^\mu\,d\lambda .
 \label{action}
\end{equation}
After variation one may use proper time $\tau$, with $u^\mu=dx^\mu/d\tau$ and $g(u,u)=c^2$. We adopt the standard contraction convention $(i_uF)(Y)=F(u,Y)$. The equation of motion is
\begin{equation}
 m(\nabla_u u)^\flat=q\,i_uF .
 \label{lorentz}
\end{equation}

A sign that is important for what follows comes directly from the first term of Eq.~\eqref{action}. The canonical momentum one-form associated with the parameterized worldline action is
\begin{equation}
 \boxed{P=-m u^\flat+qA.}
 \label{canonicalP}
\end{equation}
Its spatial components reproduce the usual coordinate-time canonical momentum because $u_i=-\gamma v_i$ with the present metric signature.

For any vector field $X$,
\begin{equation}
 \frac{d}{d\tau}\big[m g(u,X)\big]
 =m g(\nabla_u u,X)+m g(u,\nabla_uX).
 \label{metricderivative}
\end{equation}
If $X$ is Killing, the second term vanishes. Using Eq.~\eqref{lorentz} and Cartan's identity
$\mathcal L_XA=i_XF+d[A(X)]$, one obtains
\begin{align}
 \frac{d}{d\tau}P(X)
 &=-qF(u,X)+q\,u[A(X)] \notag\\
 &=q(\mathcal L_XA)(u).
 \label{PXderivative}
\end{align}
Condition~\eqref{symmetry} therefore gives
\begin{equation}
 \boxed{C_X=-m g(u,X)+qA(X)-q\chi_X,\qquad
 \frac{dC_X}{d\tau}=0.}
 \label{CX}
\end{equation}

Gauge invariance follows directly. Under $A'=A+d\Lambda$,
\begin{align}
 \mathcal L_XA'&=\mathcal L_XA+\mathcal L_Xd\Lambda \notag\\
 &=d\chi_X+d[X(\Lambda)]
 =d\big[\chi_X+X(\Lambda)\big],
 \label{gaugechi}
\end{align}
so, up to an irrelevant additive constant,
\begin{equation}
 \chi_X'=\chi_X+X(\Lambda).
 \label{chitransform}
\end{equation}
Since $P'(X)=P(X)+qX(\Lambda)$,
\begin{equation}
 C_X'=P'(X)-q\chi_X'=C_X.
 \label{gaugeinvariance}
\end{equation}
Thus $P(X)$ is a gauge-dependent canonical representative, whereas $C_X$ is the gauge-invariant symmetry charge. The special condition $\mathcal L_XA=0$ corresponds to a gauge in which the compensating function can be chosen constant.

In Minkowski spacetime translations and Lorentz transformations are obtained from the same construction. For $X=a^\mu\partial_\mu$, Eq.~\eqref{CX} gives the charge associated with the corresponding translation. For $X^\mu=\omega^\mu{}_{\nu}x^\nu$, with $\omega_{\mu\nu}=-\omega_{\nu\mu}$, it generates rotational and boost charges. In the absence of electromagnetic coupling, Eq.~\eqref{CX} is equivalent, up to the conventional overall sign associated with the chosen generator, to the usual orbital Poincar\'e generators. For the interacting problem the combination in Eq.~\eqref{CX}, rather than $P(X)$ alone, is the gauge-invariant object.

\section{Nonrelativistic quasi-invariance}
In an inertial frame Eq.~\eqref{action} becomes
\begin{equation}
 L=-mc^2\sqrt{1-v^2/c^2}-q\phi+q\mathbf v\cdot\mathbf A,
 \label{NRL}
\end{equation}
with canonical three-momentum
\begin{equation}
 \boldsymbol\Pi=\frac{\partial L}{\partial\mathbf v}
 =\gamma m\mathbf v+q\mathbf A.
 \label{Pi}
\end{equation}
Writing
\begin{equation}
 U(\mathbf r,\mathbf v,t)=q\phi-q\mathbf v\cdot\mathbf A,
 \label{Uem}
\end{equation}
we have $\partial U/\partial\mathbf v=-q\mathbf A$ and therefore
\begin{equation}
 \boldsymbol\Pi=m\mathbf v-\frac{\partial U}{\partial\mathbf v}+O(c^{-2}).
 \label{Pivel}
\end{equation}

To preserve the full content of Eq.~\eqref{CX}, however, strict invariance is not sufficient. Consider an infinitesimal point transformation
\begin{equation}
 \delta t=\epsilon\xi,\qquad
 \delta\mathbf r_i=\epsilon\boldsymbol\eta_i,
 \label{pointtrans}
\end{equation}
for which the Lagrangian is quasi-invariant,
\begin{equation}
 \delta L+L\frac{d(\delta t)}{dt}
 =\epsilon\frac{dB_X}{dt}.
 \label{quasiNR}
\end{equation}
Noether's theorem then gives the conserved generator charge
\begin{equation}
 \boxed{
 C_X^{\rm NR}=
 \sum_i\mathbf p_i^{\rm can}\cdot
 (\boldsymbol\eta_i-\mathbf v_i\xi)
 +\xi L-B_X,
 }
 \label{CNR}
\end{equation}
where $\mathbf p_i^{\rm can}=\partial L/\partial\mathbf v_i$. For the electromagnetic coupling inherited from Eq.~\eqref{symmetry},
\begin{equation}
 B_X=\sum_i q_i\chi_X(\mathbf r_i,t),
 \label{BX}
\end{equation}
with the appropriate compensating function for the generator. Equation~\eqref{CNR} is therefore the coordinate-time Noether--Bessel-Hagen counterpart of Eq.~\eqref{CX}. The apparent sign of the energy charge depends only on whether one uses the time-translation generator or its negative; below we define the physical energy with its conventional positive sign.

For a spatial translation by a constant vector $\mathbf a$, $\xi=0$ and $\boldsymbol\eta_i=\mathbf a$. If $B_{\mathbf a}=\mathbf a\cdot\mathbf b$, Eq.~\eqref{CNR} becomes $C_{\mathbf a}^{\rm NR}=\mathbf a\cdot\mathbf P_{\rm NBH}$, with
\begin{equation}
 \boxed{
 \mathbf P_{\rm NBH}=\sum_i\frac{\partial L}{\partial\mathbf v_i}-\mathbf b.
 }
 \label{PNBH}
\end{equation}
For an infinitesimal rotation $\boldsymbol\omega$, $\boldsymbol\eta_i=\boldsymbol\omega\times\mathbf r_i$. If $B_{\boldsymbol\omega}=\boldsymbol\omega\cdot\mathbf b_J$, then
\begin{equation}
 \boxed{
 \mathbf J_{\rm NBH}=\sum_i\mathbf r_i\times
 \frac{\partial L}{\partial\mathbf v_i}-\mathbf b_J.
 }
 \label{JNBH}
\end{equation}
For a time translation, $\xi=1$ and $\boldsymbol\eta_i=0$. Equation~\eqref{CNR} gives the negative of the conventional energy; hence the conserved physical energy is
\begin{equation}
 \boxed{
 E_{\rm NBH}=\sum_i\mathbf v_i\cdot
 \frac{\partial L}{\partial\mathbf v_i}-L+B_t,
 }
 \label{ENBH}
\end{equation}
where $B_t$ is defined by $\partial L/\partial t=dB_t/dt$. When the chosen gauge is strictly invariant, $B_{\mathbf a}=B_{\boldsymbol\omega}=B_t=0$, and Eqs.~\eqref{PNBH}--\eqref{ENBH} reduce to the familiar generalized constants. In particular, for $L=T-U$,
\begin{align}
 \mathbf P&=\sum_i\left(\mathbf p_i-\frac{\partial U}{\partial\mathbf v_i}\right),\label{Pstrict}\\
 \mathbf J&=\sum_i\mathbf r_i\times
 \left(\mathbf p_i-\frac{\partial U}{\partial\mathbf v_i}\right),\label{Jstrict}\\
 E&=T+U-\sum_i\mathbf v_i\cdot\frac{\partial U}{\partial\mathbf v_i}.
 \label{Estrict}
\end{align}
Thus the earlier velocity-dependent formulas are the strict-invariance representatives of the more general quasi-invariant charges, not substitutes for them.

For the electromagnetic interaction~\eqref{Uem}, the velocity-linear term cancels from Eq.~\eqref{Estrict}. The exact Legendre transform of Eq.~\eqref{NRL} in a stationary gauge gives
\begin{equation}
 E=\gamma mc^2+q\phi,
 \label{exactenergy}
\end{equation}
and hence $E-mc^2=mv^2/2+q\phi+O(c^{-2})$. More generally, if $U=U_0+U_1+U_2$, with $U_n$ homogeneous of degree $n$ in all velocities, Euler's theorem yields
\begin{equation}
 E=T+U_0-U_2
 \label{homogeneousenergy}
\end{equation}
for a strictly time-invariant representative.

A uniform magnetic field makes the boundary term explicit. Let $\mathbf B$ be constant and choose the symmetric gauge
\begin{equation}
 \mathbf A=\frac12\mathbf B\times\mathbf r.
 \label{symgauge}
\end{equation}
For a translation generated by $\mathbf a$,
\begin{equation}
 \delta\mathbf A=(\mathbf a\cdot\nabla)\mathbf A
 =\frac12\mathbf B\times\mathbf a=\nabla\chi_{\mathbf a},
 \qquad
 \chi_{\mathbf a}=\frac12(\mathbf B\times\mathbf a)\cdot\mathbf r.
 \label{uniformBchi}
\end{equation}
Therefore $B_{\mathbf a}=q\chi_{\mathbf a}$ and
\begin{align}
 C_{\mathbf a}^{\rm NR}
 &=\mathbf a\cdot(m\mathbf v+q\mathbf A)-q\chi_{\mathbf a}\notag\\
 &=\mathbf a\cdot\big(m\mathbf v+q\mathbf B\times\mathbf r\big).
 \label{pseudomomentum}
\end{align}
The conserved pseudomomentum is consequently
\begin{equation}
 \boxed{\mathbf K=m\mathbf v+q\mathbf B\times\mathbf r,}
 \label{K}
\end{equation}
which satisfies $d\mathbf K/dt=0$ by the Lorentz equation. This example shows directly why the quasi-invariance term cannot be dropped: the physical magnetic field is translation invariant although the chosen potential is not.

\section{Closed system and Darwin reduction}
When $A$ is dynamical, the globally conserved quantity belongs to the combined matter--field system. In differential-form notation the free electromagnetic action is
\begin{equation}
 S_{\rm em}=-\frac{1}{2\mu_0}\int_{\mathcal M}F\wedge *F.
 \label{Sem}
\end{equation}
The field equations imply
\begin{equation}
 \nabla_\mu T_{\rm total}^{\mu\nu}=0,
 \qquad
 T_{\rm total}^{\mu\nu}=T_{\rm matter}^{\mu\nu}+T_{\rm em}^{\mu\nu},
 \label{Tcons}
\end{equation}
and a Killing vector defines
\begin{equation}
 Q_X[\Sigma]=\int_\Sigma T_{\rm total}^{\mu\nu}X_\nu\,d\Sigma_\mu.
 \label{QX}
\end{equation}
With signature $(+,-,-,-)$,
\begin{equation}
 T_{\rm em}^{\mu\nu}=\frac{1}{\mu_0}
 \left(-F^{\mu\alpha}F^\nu{}_{\alpha}
 +\frac14g^{\mu\nu}F_{\alpha\beta}F^{\alpha\beta}\right).
 \label{Tem}
\end{equation}
Thus $C_X$ and $Q_X$ answer different questions: the former concerns a particle moving in a symmetric prescribed background, whereas the latter includes the energy and momentum carried by matter and field~\cite{Blaschke2021}.

The Darwin theory supplies the conservative low-velocity many-particle realization after field elimination. Through order $c^{-2}$,
\begin{align}
 L_D={}&\sum_i\left(\frac12m_iv_i^2+\frac{m_iv_i^4}{8c^2}\right)
 -\sum_{i<j}\frac{q_iq_j}{4\pi\epsilon_0r_{ij}}\notag\\
 &+\sum_{i<j}\frac{q_iq_j}{8\pi\epsilon_0c^2r_{ij}}
 \left[\mathbf v_i\cdot\mathbf v_j
 +(\mathbf v_i\cdot\hat{\mathbf r}_{ij})
  (\mathbf v_j\cdot\hat{\mathbf r}_{ij})\right]
 +O(c^{-4}),
 \label{DarwinL}
\end{align}
where $\mathbf r_{ij}=\mathbf r_i-\mathbf r_j$ and
$\hat{\mathbf r}_{ij}=\mathbf r_{ij}/r_{ij}$~\cite{Darwin1920,deCastro2002}. In Coulomb gauge the associated Darwin vector potential at particle $i$ is
\begin{equation}
 \mathbf A_D(\mathbf r_i)=\sum_{j\ne i}
 \frac{q_j}{8\pi\epsilon_0c^2r_{ij}}
 \left[\mathbf v_j+(\mathbf v_j\cdot\hat{\mathbf r}_{ij})
 \hat{\mathbf r}_{ij}\right].
 \label{AD}
\end{equation}
Differentiation of Eq.~\eqref{DarwinL} gives
\begin{equation}
 \boldsymbol\Pi_i^D=m_i\mathbf v_i+
 \frac{m_iv_i^2}{2c^2}\mathbf v_i+q_i\mathbf A_D(\mathbf r_i)+O(c^{-4}).
 \label{PiD}
\end{equation}
Because the field-eliminated Darwin Lagrangian is strictly invariant under common translations and rotations, its boundary terms vanish for these symmetries, and
\begin{equation}
 \boxed{
 \mathbf P_D=\sum_i\boldsymbol\Pi_i^D,
 \qquad
 \mathbf J_D=\sum_i\mathbf r_i\times\boldsymbol\Pi_i^D.
 }
 \label{PJD}
\end{equation}
The Legendre transform gives
\begin{align}
 E_D={}&\sum_i\left(\frac12m_iv_i^2+\frac{3m_iv_i^4}{8c^2}\right)
 +\sum_{i<j}\frac{q_iq_j}{4\pi\epsilon_0r_{ij}}\notag\\
 &+\sum_{i<j}\frac{q_iq_j}{8\pi\epsilon_0c^2r_{ij}}
 \left[\mathbf v_i\cdot\mathbf v_j
 +(\mathbf v_i\cdot\hat{\mathbf r}_{ij})
  (\mathbf v_j\cdot\hat{\mathbf r}_{ij})\right]
 +O(c^{-4}).
 \label{ED}
\end{align}

The correspondence can now be stated without suppressing the boundary term:
\begin{equation}
 \boxed{
 C_X\ \xrightarrow[c\to\infty]{}\ C_X^{\rm NR}
 \ \xrightarrow[\text{field elimination}]{O(c^{-2})}\ C_X^D,
 }
 \label{hierarchy}
\end{equation}
where, for the elementary generators,
\begin{equation}
 C_X^{\rm NR}\ \leftrightarrow\
 \{\mathbf P_{\rm NBH},\mathbf J_{\rm NBH},E_{\rm NBH}\},
 \qquad
 C_X^D\ \leftrightarrow\
 \{\mathbf P_D,\mathbf J_D,E_D\}.
 \label{hierarchycharges}
\end{equation}
The first arrow is a low-velocity reduction of the same quasi-invariant symmetry construction. The second is different in character: it is a conservative elimination of electromagnetic field degrees of freedom through Darwin order.

The exact worldline construction is not restricted to stationary backgrounds: time-dependent electromagnetic fields may admit a conserved $C_X$ whenever Eq.~\eqref{symmetry} holds. By contrast, the Darwin approximation is quasistatic and neglects radiation. Radiation reaction therefore lies outside the $O(c^{-2})$ conservative hierarchy; when radiation is relevant, the corresponding field energy-momentum must be retained. An ideal point charge also has a singular self-field, so a fully self-consistent treatment requires a self-force prescription or a controlled effective description~\cite{Gralla2010,Hsiang2022}. Finally, global charges such as Eq.~\eqref{QX} require the relevant spacetime symmetry, whereas Eq.~\eqref{Tcons} remains a local statement even when no global Killing vector exists.

\section{Conclusion}
A spacetime symmetry of a charged particle need preserve the electromagnetic potential only up to gauge. With the worldline action and signature used here, the canonical one-form is $P=-mu^\flat+qA$, and the gauge-invariant Noether--Bessel-Hagen charge is $C_X=P(X)-q\chi_X$. Its gauge invariance follows because the transformation of $A(X)$ is accompanied by $\chi_X\to\chi_X+X(\Lambda)$.

The main result is the explicit reduction map in Eq.~\eqref{hierarchy}. The nonrelativistic stage is now formulated at the same level of quasi-invariance as the covariant theory: the boundary function $B_X$ is retained in Eq.~\eqref{CNR}, and the familiar generalized momentum, angular momentum and energy appear only as strict-invariance representatives when $B_X=0$. The uniform-field example verifies the construction directly and produces the conserved pseudomomentum from a potential that is invariant only up to gauge. The Darwin theory then realizes the same symmetry logic after conservative field elimination, while the closed particle--field system is described instead by the stress-energy charge $Q_X$. The novelty is therefore the explicit, gauge-consistent reduction connecting these established descriptions, not a claim that any one of their individual conservation laws is new.


\begin{thebibliography}{99}
\bibitem{Noether1918}
  Noether E.
  \textit{Nachr. Ges. Wiss. G\"ottingen, Math.-Phys. Kl.}, 235 (1918).
\bibitem{BesselHagen1921}
  Bessel-Hagen E.
  \textit{Math. Ann.} \textbf{84}, 258 (1921).
\bibitem{deCastro2002}
  de Castro A. S., Marchesetti E. L. and Feldt A.
  \textit{Rev. Bras. Ens. F\'is.} \textbf{24}, 278 (2002).
\bibitem{Darwin1920}
  Darwin C. G.
  \textit{Philos. Mag.} \textbf{39}, 537 (1920).
\bibitem{Blaschke2021}
  Blaschke D. N. and Gieres F.
  \textit{Nucl. Phys. B} \textbf{965}, 115366 (2021).
\bibitem{Klinger2024}
  Klinger M. S., Leigh R. G. and Pai P.-C.
  \textit{Nucl. Phys. B} \textbf{998}, 116404 (2024).
\bibitem{Gralla2010}
  Gralla S. E.
  \textit{Phys. Rev. D} \textbf{81}, 084060 (2010).
\bibitem{Hsiang2022}
  Hsiang J.-T. and Hu B.-L.
  \textit{Phys. Rev. D} \textbf{106}, 125018 (2022).
\end{thebibliography}
\end{document}